# Improving the Technical Aspects of Software Testing in Enterprises

Tim A. Majchrzak
Department of Information Systems
University of Münster
Münster, Germany
tima@ercis.de

***Abstract*—Many software developments projects fail due to quality problems. Software testing enables the creation of high quality software products. Since it is a cumbersome and expensive task, and often hard to manage, both its technical background and its organizational implementation have to be well founded. We worked with regional companies that develop software in order to learn about their distinct weaknesses and strengths with regard to testing. Analyzing and comparing the strengths, we derived best practices. In this paper we explain the project's background and sketch the design science research methodology used. We then introduce a graphical categorization framework that helps companies in judging the applicability of recommendations. Eventually, we present details on five recommendations for technical aspects of testing. For each recommendation we give implementation advice based on the categorization framework.**

***Keywords: Software testing, testing, software quality, design science, IT alignment, process optimization, technical aspects***

## I. INTRODUCTION

Striving for improved software quality is no new emergence. The idea to optimize technical aspects respectively to use technology to achieve this aim is known for decades. Unsurprisingly, the term *software engineering* [28] has been coined in the 1960s and the *software crisis* is known—and unfortunately still lasting—since the 1970s [8].

Especially large-scale projects that end in disasters nurture the public's picture of unreliable software. An example is the NASA Mars Climate Orbiter, which crashed because metric and imperial units were mixed in a software subsystem [27]. The miscalculation leading to the crash would most likely have been detected by detailed software testing. Unfortunately, there are many other examples of failed major projects that have similar root causes: inscrutable, ill-designed or not exhaustively tested software [17].

Despite the widely perceived disasters, the *main* problem is failure of everyday projects [6][10]. Even after decades of research, no *silver bullet* has been found and many problems remain unresolved [4]. Complexity of software obviously increases faster than methods to control it are developed [16]. As a consequence, problems of varying severity can be found in projects in any industrial sector and for any kind of software developed. But not all software development projects fail; in fact, many companies produce software systems of notable quality. We propose to study effectual development to discover best practices for reaching quality especially with regard to testing. In combination with the processes and techniques for the development of software, software testing is the foundation of software quality [17].

To better support businesses with results from academic research, a combination of research in information systems and software engineering is a feasible approach [21]. We undertook a project with regional enterprises and tried to learn what makes software development projects *successful*. After identifying the companies' status quo [21], we analyzed the myriad of observations we made and the experiences the project's participants shared with us. Eventually, we derived a set of novel best practices.

It appears to be easy to say how software development should be done. But although techniques are described in the literature and there is knowledge about successful development, this knowledge has not necessarily been transferred into business reality. Some of the best practices we found have been denoted earlier e.g. in different contexts or with different prerequisites. However, adopting them seems to be very challenging. We thus give details on how to implement the recommendations and which conditions have to be met. We also name related work for each recommendation *rather* than discussing them in a section of their own. Best practices presented in this work have a technical focus; suggestions for the organizational embedding of testing can be found in [20].

This paper is organized as follows. Section II introduces the project's background. We sketch our research methodology in Section III. A framework for categorization is explained in Section IV. Five effective technical recommendations are discussed in Section V. A conclusion is drawn in Section VI and future work is highlighted in Section VII.

## II. BACKGROUND

Münster is located in North Rhine-Westphalia, Germany. In the city and its surrounding region a lot of IT-based companies have been sited. Most of them are medium-sized and specialize on software development. Some larger companies with far over 1.000 employees do not develop software for customers; as financial service providers their individually developed software enables their business processes. The number of their developers exceeds the number of employees most of the smaller companies have in total.

All companies are members of the local chamber of commerce which supports the *Institut für Angewandte Informatik* (IAI – Institute of Applied Informatics). The IAI is hosted by the University of Münster and based on the work of both computer scientists and economists. Projects undertaken by the IAI are run by academics that seek both research progress and mean to support the local industry.

By frequent exchange with companies the IAI learned about their dissatisfaction with software testing. While most companies were ambitious to improve the quality of the software they developed and to cut down costs for testing, they did not know how to achieve this. Additionally, many enterprises lack the time to try out new technologies or to evaluate changes to their processes. However, the companies were not economically endangered and apparently developed software of quality. Thus, two conclusions could be drawn. Not a single company has a *perfect* testing process. All of them face a number of testing related problems. Nevertheless, each company has developed distinct strengths that help it in creating *good* software products.

Based on these observations the IAI project to improve software testing was initiated. Two main purposes were set: Firstly, the status quo of software testing in the regional enterprises was to be brought to light. Secondly, successful strategies used by the companies were to be identified and aggregated to best practices. In this work we present five major best practices that change or influence the technical way of software testing or the technology used.

From the exchange with the enterprises and due to the diversity of software developed as well as the differences in culture in each company, we expected strengths to be complementary. Hence, it could be estimated not only to find a few known methods for successful development but a plethora of promising attempts to increase software quality and to optimize processes.

Diversity is both a blessing and a curse. It helps to identify best practices that form recommendations unknown to most companies and therefore highly beneficial to them. At the same time, prerequisites have to be met so that a recommendation can be adopted at all. Consequently, a framework is needed to support companies in choosing which recommendation to implement. The framework is described in Section IV and used for each recommendation in Section V.

## III. Research Methodology

The project was meant to combine scientific rigor with relevance and efficiency as demanded by businesses. We decided for a methodology based on *design science* which "addresses important unsolved problems in unique or innovative ways or solves problems in more effective or efficient ways" [15]. It of course is impossible to describe the *perfect* testing process or to offer a general description on *how to* test software. However, we searched for a larger number of satisfactory solutions that address typical problems. Finding such *satisficing* [31] solutions helps enterprises even though not all possible problems can be addressed.

Since we wanted to learn about problems from the point of view of the participating companies, we decided for a qualitative approach [26]. Best practices can hardly be found with a simple questionnaire. Thus, we conducted semi-structured expert interviews. Using only a rough guideline for the interviews [19], we were able to learn about how testing is done in the companies. As the interviews developed, distinct weaknesses and strengths could be identified as well as common problems and successful strategies discussed with the participants. The data gained in each interview is far too verbose to be published as such. But each of it forms a kind of case study [36] which greatly aids further analysis.

Recommendations derived from the discussion with the interview partners are meant to complement the literature. Even comprehensive work on software testing processes [2] or quality improvements [19] does not cover all problems typically faced by practitioners. Some ideas published also do not seem to be directly accessible to practitioners. Along with literature, that promotes result-driven testing [13], we want to help closing this gap. Technical aspects as depicted in this work should be given special attention. If conducting IT research, it should be kept in mind that *information technology* is studied [25]—even if organizational aspects are likewise important.

A quantitative analysis would augment the qualitative approach. Without quantitative data it is hard to *prove* that a recommendation is effective and efficient. However, deducing best practices is a first step and was very laborious; verifying results was identified as a further step (see Section VII).

The course of action we took can be sketched as follows. We began by contacting IAI supporting companies and by identifying staff for the interviews. Both managers and technically skilled employees were chosen. In a second step we interviewed the participants. While there usually was only one longer interview done with smaller companies, medium-sized and larger companies were visited more than once. We were able to address both organizational and technical issues with the respective experts. In the interviews we tried to identify *who* is responsible for testing, *when* it is done, *what* is included in tests (graphical user interface, interfaces to other systems, etc.), *which* methods are used and *how* testing is generally done. We also tried to learn about the usage of *testing tools* [23].

After discovering the status quo, we discussed general problems met and successful strategies found. This included evaluating which improvements the participating companies desired. Eventually, potential best practices were discussed with them. This part of the interview was the most open one. A lot of ideas were exchanged and many interesting approaches considered.

The third step was to analyze the results and to aggregate data. As it is of high interest to the regional companies, an overview of the status quo has been drawn. For reasons of space and scope it is not included in this paper but can be found in [21]. By finding interdependencies as well as aligning and judging best practices identified by the participants, we extracted recommendations. Of course, particularities of the companies' situations were taken into account. This lead to the construction of a framework (see Section IV) that describes the conditions under that a recommendation applies.

## IV. FRAMEWORK

Recommendations for a topic as complex and intertwined on various levels as software development require a sophisticated categorization. Their full value can only be accessed if it is known *how* to use them and which prerequisites have to be met. Besides, support on deciding which best practices are most applicable for the own business is advisable. We thus use the framework first described in [20] to classify recommendations.

The *level of demand* shows how great the organizational change required to adopt a recommendation is. *Basic* recommendations should be adopted by any company. If recommendations are not only basic hints but require considerable effort to be implemented, they are considered to be *advanced*. Eventually, *target states* are ideals that cannot be reached unlabored. In fact, they are guidelines on what level of perfection can be reached and require a process of continuous optimization. However, the benefits of an actual implementation will be immense in the long run.

It is important to consider the *project size*. *Small-sized* projects commonly have a single team that does development and testing. For *medium-sized* projects these tasks are undertaken by at least two teams. *Large* projects can comprise hundreds of employees and include general departments that contribute to it. If thinking about a recommendation, not just the sole number of employees that participate in it should be taken into account. In fact, the typical size of projects as well as their character should be kept in mind.

Another important determinant is the *kind of software* developed. Based on contracts, *individual software* is developed for a single customer. Usually, there is close contact to the principal. Standard or *mass market software* often is developed over a long period of time. This makes regression testing important. Many recommendations can be applied to both kinds of software.

Similarly, the *number of releases* of a software product has to be taken into consideration. It is differentiated between *one*, *several* and *regular* releases whereas regular means that there will be releases for some month or years.

The fifth determinant distinguishes between the *phases* (or *stages*) of testing. It is divided into the phases of *component test*, *integration test*, *system test* and *acceptance test* that also can be found in the literature [35].

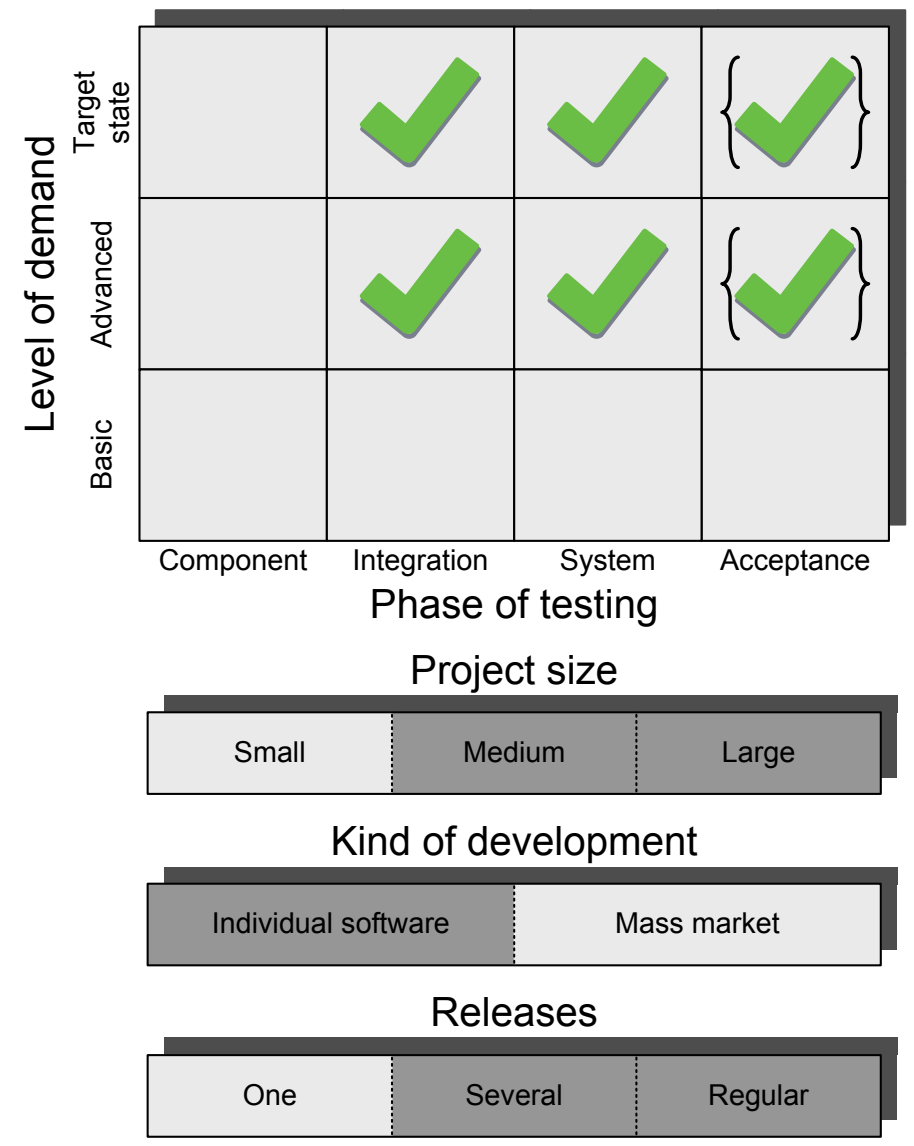

Figure 1. Exemplary use of the framework

As represented in Figure 1, the level of demand and the phase of development are used to set up a matrix. A tick indicates that a recommendation is meant to be beneficial for the depicted phase and level. Ticks might be shown in brackets which indicate that benefits will be observable but might be less pronounced than for other phases and levels. The three other determinants are shown as bars. A shade of (dark) gray means that a recommendation applies under the specified conditions. Fading indicates that adoption of the recommendation should be considered if the depicted determinant is met. Recommendations still require more detail so that companies can judge them. However, the framework can be used to get a quick overview of the main prerequisites for it.

Please consider Figure 1 for clarification:

- The recommendation requires advanced effort. It is possible to be extended in order to mark a target state in which beneficial effects will be much stronger.
- Implementing it especially aids integration and system testing. Positive effects are also likely to be observed for acceptance testing.
- The recommendation is meant to be adopted for at least medium-sized projects and it aims at individually developed software.
- It aims at individually developed software. Theoretically, there could be a fading of the gray shade into the box for mass market software. This would mean that it would also benefit while the main focus was individual software.
- For full effect, there should be a greater number or regular releases of the software developed.

## V. Technical Recommendations

The following sections present five recommendations for the optimization of technical aspects of software testing. Their order reflects the implementation complexity.

### A. State-of-the-art Development Environment

The first recommendation is pretty straightforward. We encourage using the latest development environments available, particularly integrated development environments (IDE) that are customizable and support plug-ins. They offer magnificent opportunities to increase the quality of the developed software. Using the latest IDEs is especially appealing since development software is used anyway and many of these products or at least plug-ins for them are *free*.

Admittedly, using an IDE is not only about testing. But the support it offers significantly helps to increase development quality. If the developer is aided in his routine work, testers do not have to struggle with defects in programs that originated in unthoughtfulness. Testers can then concentrate on finding actual bugs e.g. in algorithms. Consequently, this recommendation is a testing best practice even though parts of it do not *directly* deal with testing; they have a noticeable indirect impact.

Unlike expectation, companies do not necessarily use state-of-the-art IDEs. It is common to do so for individual developers in small enterprises. However, once the choice of development tools is not solely based on developers' discretion but there are general guidelines or even mandatory directives, tools that do not offer as much functionality as would be possible are used. This is especially true for situations in which developer PCs are centrally set-up by IT organization staff rather than by the developers themselves. Changing development tools could not be possible since tools for cooperative work or versioning, or software to access corporate-wide storage systems or resource pools might not be exchangeable.

Some of the participants drew a picture of the way their development is supported by the tools used that reminded us of the 1980s. There was no kind of *syntax highlighting* and no built-in supportive functionality to aid the developer with coding. There was no direct access to the programming languages or library documentation; developers would look it up on the Internet or use books even for the simplest questions. And, probably worst, there was no testing and debugging support. Debugging was done by putting `print()`-statements into the code that almost arbitrarily supplied the developers with fragments (or rather shreds) of information.

Seeing how much more productive developers using modern IDEs are and how much these tools aid them in achieving high quality software, we strongly recommend using up-to-date development environments and the functionality that comes with them. This general recommendation is suitable for any company. It is extremely helpful for component testing (see Figure 2).

If IDEs are used that do not offer some of the more sophisticated functions and cannot be extended—e.g. with plug-ins—upgrading to a more recent version or another IDE is recommended. Eclipse arguably is the most widely known and one of the most powerful IDEs. It supports Java, C/C++ and (by using extensions) many other languages such as PHP. Even though a lot of functionality is built-in, there is a four-digit number of plug-ins to enhance it further (an exemplary site that lists them is [9]). To benchmark the development environment used, it is a good idea to compare it to leading IDEs. Speaking with the participants showed that some of them used IDEs that were far from offering what Eclipse or Microsoft Visual Studio (the leading tool for .NET) do. Partly the functionality does not even reach what the leaders provided years ago.

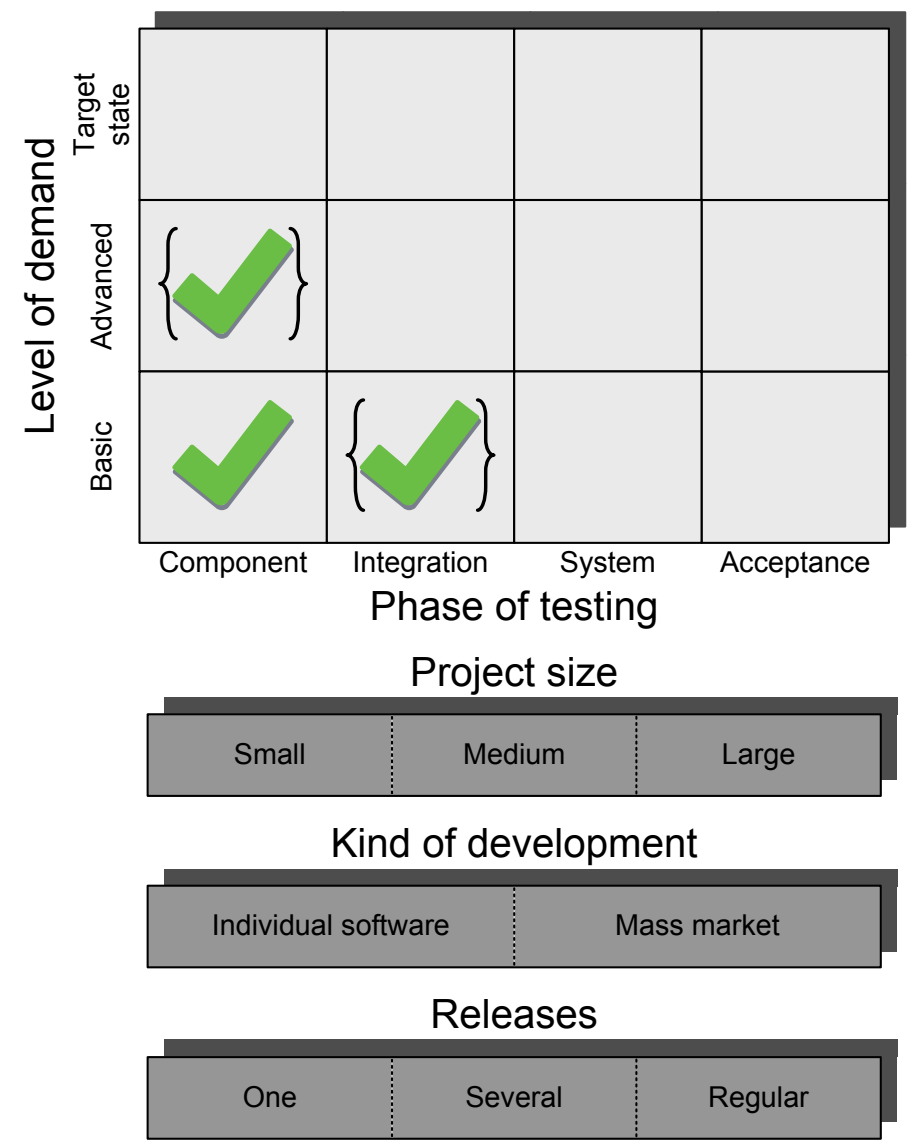

Figure 2. Classification of Development Environment

Coloring the source code to point up the syntax (syntax highlighting) [7] and automated suggestions while typing (code completion) are common. Documentation fragments can be shown directly to e.g. prevent the usage of methods marked as deprecated. Many IDEs also offer direct checking of the code so mistakes are immediately highlighted. Partial compilation can provide error information without the need to explicitly invoke the compiler. Thus, software with syntax errors will not even be tried to compile and will be fixed by the developer before they consider it to be finished.

Semantic correctness cannot be guaranteed automatically but many typical mistakes can be prevented. For example, levels of warning can be defined. We strongly encourage enabling this feature. Eclipse can for example show Java warnings by underlining code in yellow color. A variable that may take the value of `null` but is used without checking for this will be marked. Consequently, code that provokes so called `NullPointerExceptions` can be fixed. Many other mistakes can be prevented from being made. Despite an unfamiliar feeling programmers might have in the beginning, they are getting used to the warnings quickly. Superfluous warnings usually can be disabled; in Java this e.g. can be done by using so called annotations [3].

The next step is checking *code rules*. IDEs do not offer this functionality but there are tools and plug-ins available. While the above described warnings are generated by the compiler and shown by the IDE, tools for checking code rules have a logic on their own which makes them more powerful. Moreover, they are customizable and allow having corporate-wide coding standards enforced. While developers should not feel patronized, having common standards is highly recommended. Many problems arise when several developers work on the same code and probably misunderstand what their colleagues did. This is particularly problematical if developers introduced the style of their choice and then leave the company while the code they wrote has to be maintained. This can be prevented by having corporate-wide schemes and conventions. We suggest using tools or plug-ins to check compliance with general coding standards suggested by the programming language vendors (e.g. [33]), literature (e.g. [34]) and company-specific additions.

We also advocate using the debugging functionality of modern IDEs. Instead of printing out variable contents, modern *trace debuggers* visualize the complete state of a program at a point of the developer's choice. Pointers can be followed and variables modified; execution can be continued step-by-step. Visualizing graphs for control flow and data flow further aids debugging. Combined with knowledge on modern debugging techniques [11] the debugger of a state-of-the-art IDE is a tool of immense power and versatility.

To sum up, we strongly recommend using a modern IDE, even if giving up old libraries, methods, paradigms or even programming languages is a precondition. Along with this process, binding standards for formatting source code and for naming variables, methods etc. should be set up. For a better understanding how the optimal usage of a programming language can be supported by an IDE, practitioner literature such as [3][24] is recommended. There is a plethora of work on programming best practices that can be utilized to augment this recommendation.

### B. Test Case Management and Database

In small projects testing often is seen as a stateless task. Tests are done once a module is finished and found defects are corrected directly. This is repeated at the levels of integration and system testing. Unfortunately, it is inefficient and cannot be combined with a holistic view [20] of testing. Hence, we recommend using a test case management tool. It already helps medium-sized projects that have at least a couple of releases. While the later phases of testing are supported with little effort, the solution can be expanded and will be beneficial for all phases of testing (see Figure 3).

Typical functions include the compilation and categorization of test cases, ideally using a highly customizable interface that supports users with suggestions to disburden them of repetitive tasks, and setting statuses of test cases. Optionally, assignment of tasks and responsibilities can be done. A tool should further support cooperative work and offer reminders (via e-mail) for employees about assigned tasks, nearing deadlines and other important dates. Connections to the environments that run test cases are another amenity. They allow testing to be triggered automatically.

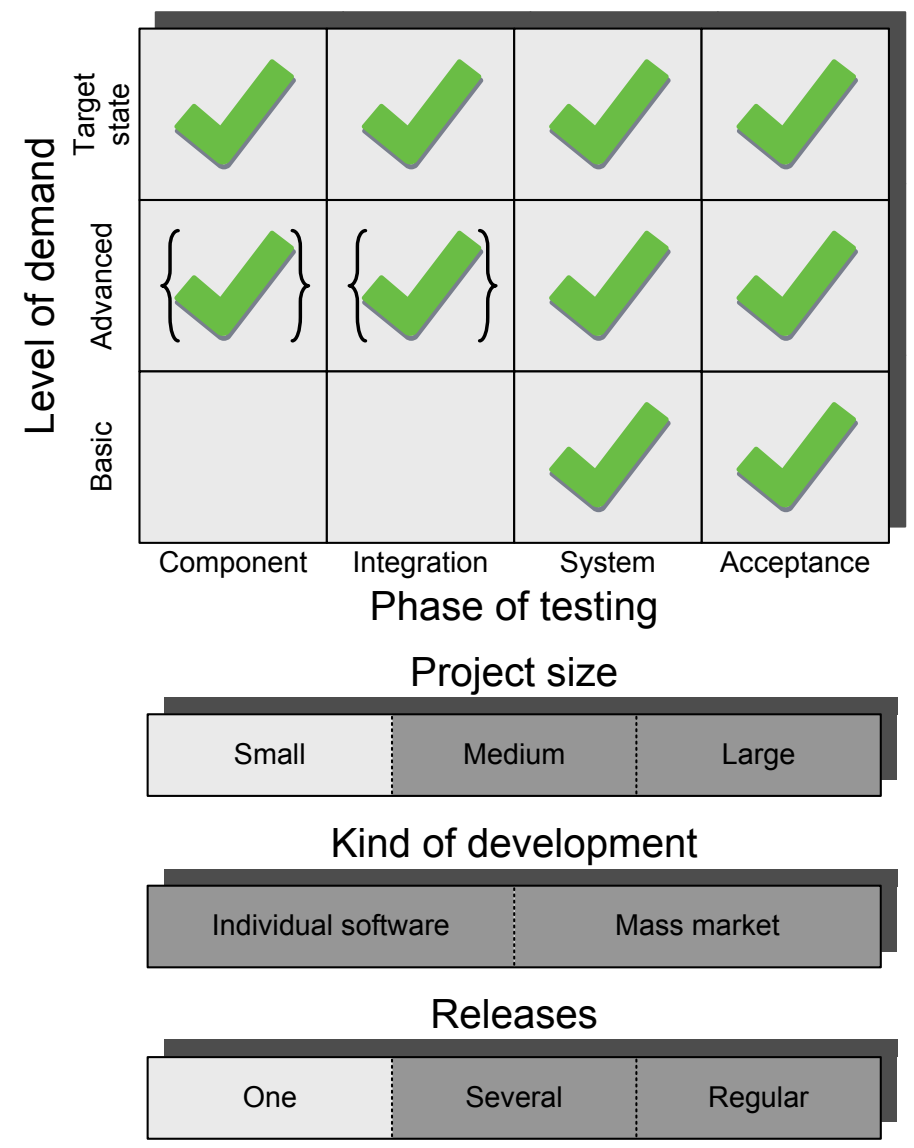

Figure 3. Classification of Test Case Management

Thus, the main purpose is formalization and structuring. Ideally, each employee knows exactly what he has to do at any time and can look up that information in a test case management tool. To a certain degree he can choose from tasks yet unassigned. When pursuing these tasks, he likely will spend his efforts with high efficiency. The tool should also be able to report a project's status which is especially helpful for large projects. The added effort for entering test cases can be minimized with intelligent help from the tool. Besides, regression testing is improved.

Despite not many facts on test case management being published, we know of one detailed work. Parveen et al. present a case study [30] on the implementation of a centralized test management using *TestDirector*, a tools by then sold by Mercury Interactive. While the study is different in context and scope, experiences are similar to our observations of the beneficial effects of test case management.

The test case management's functionality can be extended successively. Not only can it be used more precisely but additional functionality can be added. It is a good idea to include support for *requirements engineering*. Tasks can be derived from requirements and test cases can be linked to them. Should test cases fail, the employee responsible for the requirement might be able to help. Reporting can also help to find modules that have a high rate of defects which probably result from mistakes in their requirements.

Especially for products that are continuously refined, integration of a *bug tracker* is recommended. This software is used to report and manage defects (*bugs*) and therefore ideal for integration with test case management. Bug trackers can become an interface to the technical staff of the customer. A wealth of further functionality can be easily added.

Test case management is thought to be an interface between steps of processes. Erstwhile informal and hardly checkable process components are represented by it. Information is pro-vided in a structured form. The management software is used for any operative testing procedures. In fact, each testing process begins and ends with utilizing it.

Beginning with using table calculation sheets can foster the creation of an integrated system that offers interfaces to other tools. Expanding the test case management should be done step-by-step. Both a bottom-up (beginning with component tests) and a top-down approach (beginning with system tests or even acceptance tests) are possible. Media disruption should be avoided as it lowers efficiency. An example for media disruption is to write results from a test run to a piece of paper and later type the results from the paper into a tool. If a strategy for implementation is worked out in advance, a delay of projects is unlikely and a return-on-investment (ROI) should be achieved timely. While implementation details are out of scope of this paper, we strongly advise setting up a test case management.

On the first look a *test case database* appears to be equal to test case management software. While both purposes can be combined in one tool, there is a functional distinction between them. Test case management serves towards the aim of structuring and documentation. A test case database is driven technically. It is used to collect test cases in executable form and stores components such as *test stubs* and *mock objects*. The main aim is to increase the rate of test case reuse and hence to facilitate regression testing.

Test case databases are usually integrated into tools but can be implemented separately. Test cases have to be saved in a structured way and it should be easy to find and retrieve them. Ideally, the database system can directly invoke the environment test cases are coded for and run them. It is very helpful for data-driven applications if (e.g. relational) databases can be stored along with test cases. Arbitrary testing results are prevented since the database can be reset to a defined and consistent state for any test cases that requires this.

A test case database has amenities beyond the mere reuse of test cases. A good strategy for testing larger software systems is to have a defined suite of test cases and run it both for an old, correctly working version *and* the new version of the software. If results differ, defects are likely. The same applies to test databases. First the database is set to a defined state. Then the test suite is run for the old version of the software. The same is repeated for the newer version of it after the database has been reverted to the defined state. Resulting states are compared since differences hint to problems. If results are identical but the old version is known to be *buggy*, problems have apparently not been fixed. While such testing is possible manually, tool support avoids mistakes and saves much manual labor.

Test case databases are also useful if libraries are developed that are incorporated into several other systems. They can be tested even if changes were made. Changes to interfaces or defined functionality are noticed immediately without deploying the library to productive systems.

The strengths of test case databases are most apparent if regression testing is used. Consider an example: If two algorithms for the same purpose but with different runtime characteristics have to be tested, test cases have to be implemented only once. The test cases can simply be reused. It will only be needed to add more test cases if the new algorithm has an extended functionality. With a good test case management, this is even true if the second algorithm has been implemented month or years after the first one. Without such a system, the old test cases most likely have been deleted in the meantime, were lost along abandoned data, or there will be no knowledge how to use them.

We advocate both using test case management and a test case database. They are especially successful if they are integrated (see Section V.D).

### *C. Aligning Systems for Testing and Production*

Utilizing modern programming languages and paradigms for developing complex distributed applications does not only bear advantages. Developing applications on workstations but deploying them to servers or mainframes is prone to compatibility and scaling problems.

In the very beginning of programming, software only ran on the system it was developed for. For any other platform the code at least had to be adjusted. It might even have been easier to rewrite it from scratch if architectures were entirely different. Nowadays the environment used for development and testing usually differs from the one software is developed for. Moreover, at least an operating system is mediating with the hardware. In most cases virtualization hypervisors, application servers and other components form additional layers. This has a plethora of amenities. Using high level programming languages allows for the compilation of the same code for different platforms; virtual machines and other components can even offer hardware abstraction. However, the productive system often is far more powerful and not only its hardware is different but often the software is different, too.

In simple cases, differences only apply to the workstations and servers' operating systems. However, additional components such as libraries, database management systems (DBMS) or application servers are likely to be different as well. Server versions of these systems will probably not even run on workstations. Consequently, problems arise. To give an example: Java EE applications are commonly run in a sophisticated application server such as IBM *WebSphere*. Workstations often run a lightweight Apache *Tomcat*. Even though an application that runs on Tomcat should seamlessly do so on WebSphere, practice shows that unexpected behavior or crashes can be expected. This can be explained with a different interpretation of specifications, differing versions, conflicting libraries and similar issues.

We recommend aligning development and testing systems with the intended productive environment. By *alignment* we mean to reasonably adjust development and productive hardware and software while keeping the effort economically feasible. It will in most cases e.g. not be justified to buy a second mainframe system just to have a testing platform that is equal to the productive system. Nevertheless, options are often available that guarantee a high technical compatibility but are cost-effective. Exactly to find these options alignment is about.

Business/IT alignment in general is subject to lively scientific discussion [5].

Aligning systems is suggested for at least medium-sized projects with a couple of releases. It is especially useful for individually developed software and early development phases (see Figure 4). Due to our observations we even deem additional effort to align systems justified. Surprisingly, no work seems to be published on system alignment for the reasons of testing.

The more advanced a testing phase is the more alike should systems be. Compatibility problems should however be resolved as early as possible. Achieving this can be easier than thought. For example, lightweight versions are available for common server applications. This applies to the earlier WebSphere example; Tomcat should be used on the client only if the target system is Tomcat either.

Instead of installing a DBMS on the developing system, the one installed on the server can be used remotely. A separated database should be created to protect productive data from corruption. Modern servers and to an even higher degree mainframes offer virtualization that allows to completely separate instances not only of databases but of any application. Thus, testing is possible on the same machine and with the same system software that the application will eventually run on. Resource usage should be protected so that a tested application running into a deadlock or massively using resources does not endanger productive applications running in parallel.

Applications accessed by a number of parallel users require realistic testing. Problems that arise with memory usage or parallel execution can hardly be found with systematic testing. Such problems will not reveal themselves if just "trying out" the application on the testing system. An acceptable performance on the testing system cannot be assumed for the productive system even if it is more powerful. Not yet considered dependencies, growing data and similar issues can cause problems in the (far) future. Thus, testing *has* to be done under realistic conditions. Defects in parallel algorithms might only reveal themselves under certain conditions. *Race conditions* in which several threads of an application obstruct each other will probably occur on fast systems only. Reasonable conclusions about an application's performance can solely be drawn when thoroughly testing them in a productive environment.

Besides all advocating to testing under realistic conditions, we strongly advise *not* to test on productive systems without making sure that productive data cannot be modified and that the performance remains unaffected. Negative (side-) effects on productive systems would render any benefit of realistic testing useless.

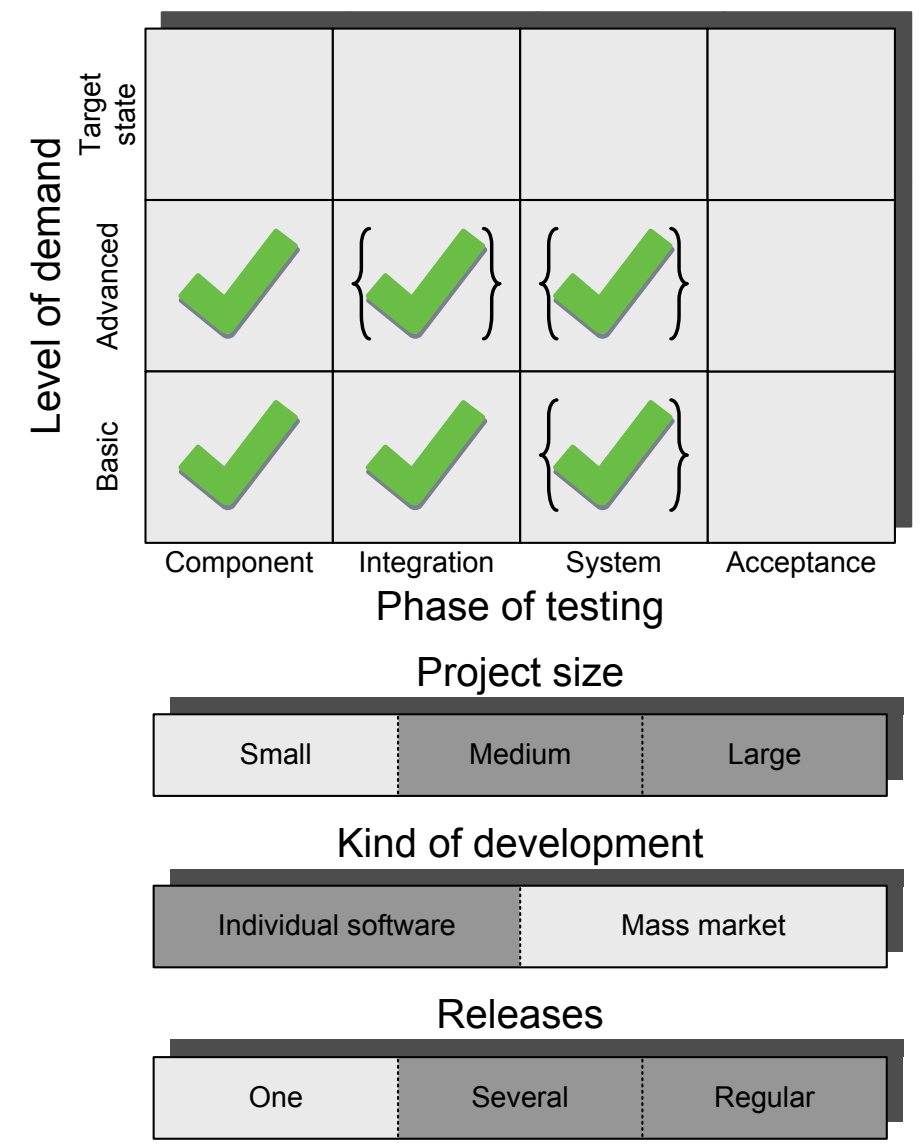

Figure 4. Classification of Aligning Systems

## D. Integration of Tools

We learned from the participants that using tools for testing is common. A general observation was that tools are hardly integrated. However, exactly this is recommended.

Testing tools are applications on their own in most cases. Common formats or defined interfaces seldom exist. Only larger tools such as *IBM Rational* products provide an interchange of data. Most participants desired the integration whereas only few of them actually had experiences with it. We recommend it for medium-sized and larger projects with at least a couple of releases. Due to the high complexity some effort is required before benefits can be observed for the phases of integration and system testing. Ultimately, amenities can be realized for all phases (see Figure 5).

Several kinds of integration are desirable. First of all, documentation systems should be linked with systems for testing. Undocumented tests are only worth a fraction of documented ones. Automatically synchronizing the results of execution with the test case management system (cf. Section V.B) disburdens testers of repetitively entering test cases. A well structured documentation as described in [16] can be achieved more easily. An improved database of testing results can also be used for statistical examination. Test managers can easily learn about running times, success rates of test cases and similar data. For regularly released software integrating the bug tracker with the management system is another option. Reported bugs can be adjusted with known defects and test cases. This decreases redundancy.

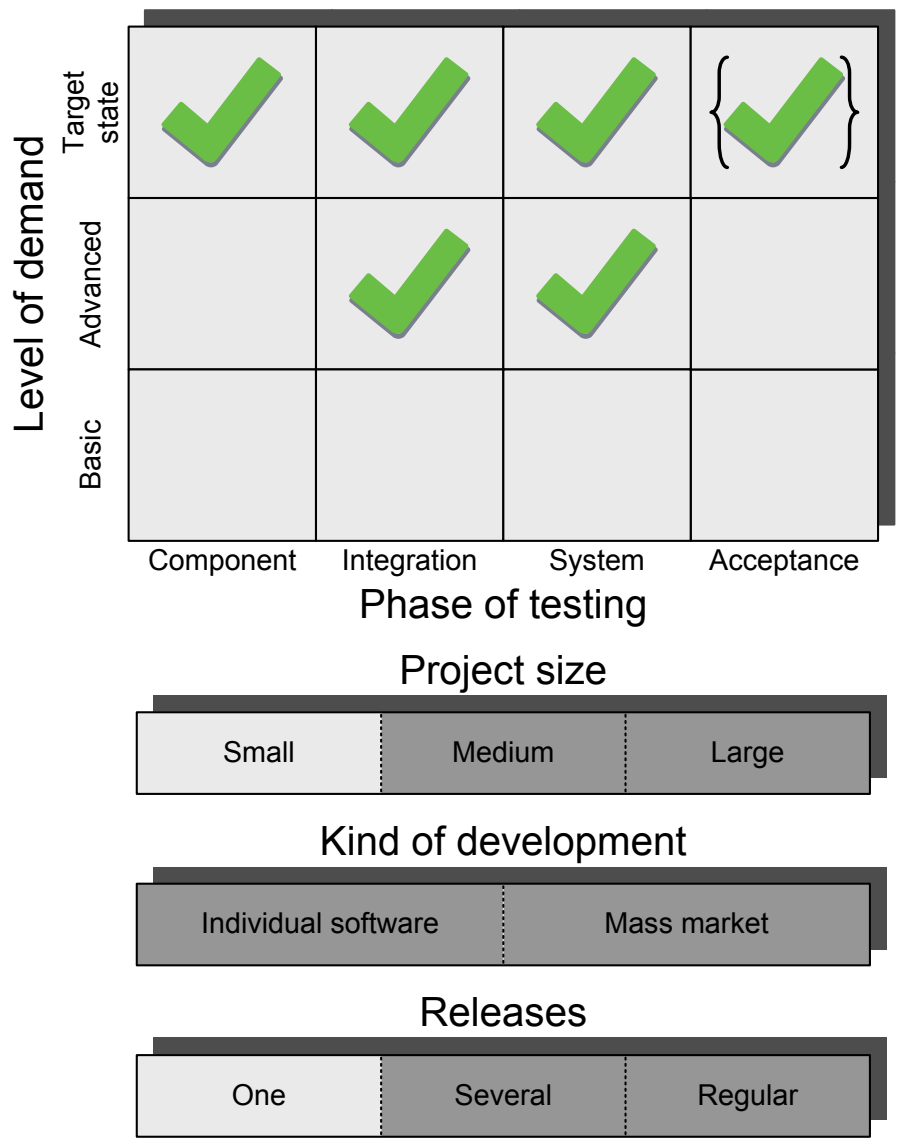

Figure 5. Classification of Integration of Tools

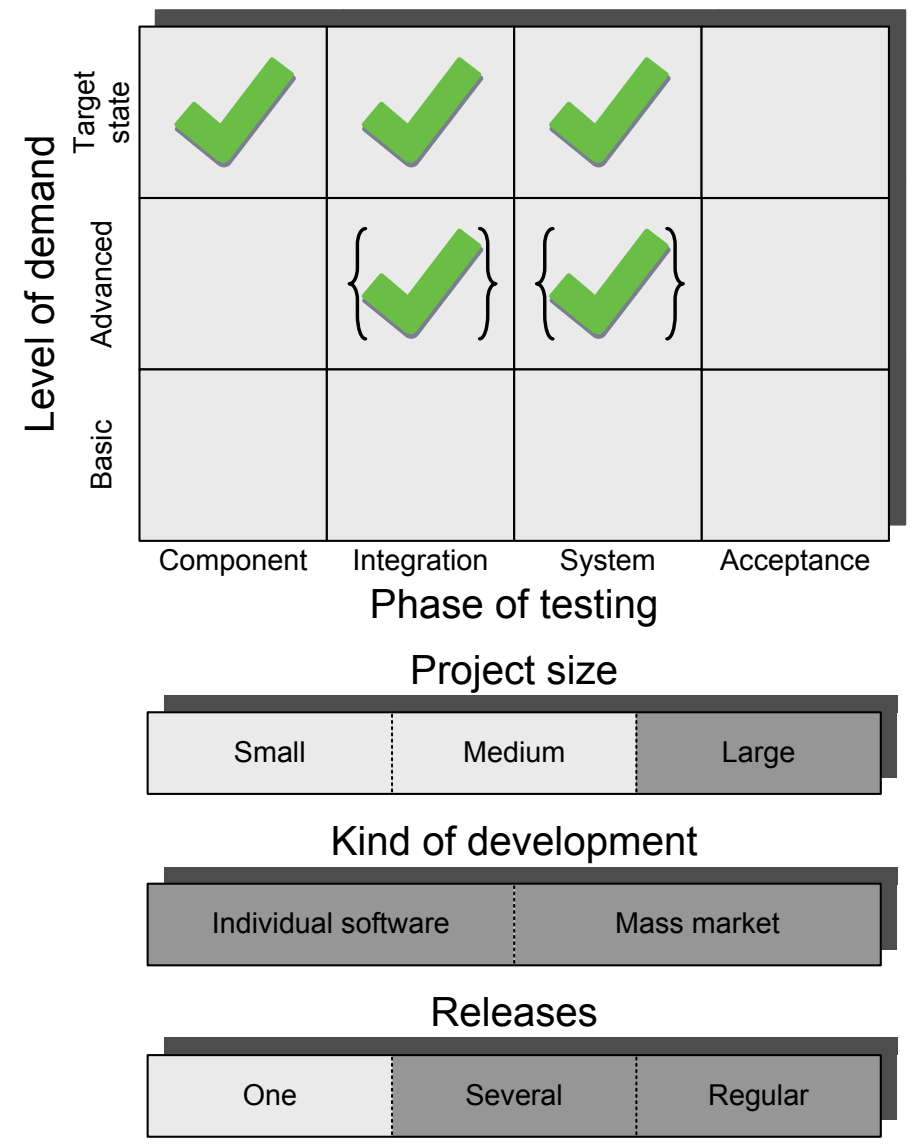

Figure 6. Classification of Customizing of Tools to Fit wih Processes

Linking systems for test case execution supports regression testing. Test cases run in earlier phases can easily be repeated. Automated synchronization again relieves testers of repetitive tasks and ineffective work. Connections to test case management systems and a test case database can make tasks economically feasible that would be too laborious if done manually.

The above ideas will seem utopian in a development landscape without integration. They should motivate alignment and encourage challenging the status quo. To our knowledge there is no exhaustive solution offered and there are no well-defined standards. Individually developing tools for integration will be unavoidable. Nevertheless, for tools newly bought integration capabilities can be checked. Even small tools for data transformation can yield dramatic reductions of manual workload. A tool for aggregating data and computing statistical reports from the test documentation can e.g. be implemented with little effort and refined continuously.

By undertaking a strategy of small refinements, integration is possible without much trouble or high costs. Growing knowledge will bolster further development. We found open source software to be convenient for integration. It can be modified to work with existing software with ease.

Full integration of tools enables new possibilities. This includes installing a *test controlling* which is used to keep an overview of the testing process and to calculate key figures [20]. The vision is an integration of systems that comprise test case management, development (project) planning, test scheduling, staff assignment, time control, task management, controlling and even a *management cockpit*.

### *E. Customizing of Tools to Fit wih Processes*

In most cases, testing tools are driven by the underlying technology. Even if they can be customized, they induce a certain way in which they have to be used. As a consequence, business processes are changed in order to fit with a tool's requirements. Without changing the processes, many tools can hardly be used. Alternatively, customizing tools is possible but very laborious. However, tools should be tailored to fit with business processes and not the other way around.

Especially companies that have *defined* testing processes pointed out, that changing processes to enable the usage of tools is a particularly bad idea. In fact, tools should be customizable in order to seamlessly integrate into the processes. Therefore, we recommend selecting tools based on their customizability. While introducing a new tool could be used to benchmark the affected business processes, well performing processes should not be changed. Customizing tools should be done in larger projects with at least several releases of a software product. The benefits will become most obvious if a company strives for a continuous optimization of its testing processes. Optimizations will be most apparent in tool-driven phases—hence, there will be hardly an effect on acceptance testing (see Figure 6).

In general, introducing new tools or installing upgrades of existing tools entails changes. They for example are caused by the implementation of additional phases of testing or the addition of new functionality. This kind of changes is both normal and desirable. Companies should try to optimize their processes, though. Adapting the course of action and procedures given by the tool should be a starting point for own considerations. Only in a small number of cases these presets will align with a company's standards. Consequently, a well-founded strategy of integrating a tool has to be found. Moreover, evaluations of the tool's performance should be scheduled. Experiences gained after using it for a while should be used for further improvements.

Implied processes are often based on technical details of tools. We learned in the underlying project of this article that only a small number of testing tools can be intuitively used.

Thus, tools should be checked for their customizability upon evaluation and selection. Steps for creating a test case should be designed to align with employees' flow of work. If the documentation, demonstration materials, or tool presentations hint to fixed and unchangeable presets, tools have to be carefully checked. It is particularly impedimental if enforced processes cannot be divided into substeps or if tools lack interfaces. A common problem would be tools for test execution that cannot be integrated with documentation software.

Adaptability and customizability can be given in several dimensions. Technically speaking, it should be possible to interrupt tests during execution in order to save intermediate results. Moreover, interfaces to import and export data are very helpful—in particular, if they can be used in real time (cf. Section V.D). With regard to the usability, a configurable interface positively affects the acceptance of a tool. The possibilities to tailor a tool should be based on its complexity. Customizing in the technical dimension (e.g. by writing scripts) is acceptable for small tools only. Ideally, tools should offer the possibility to load plug-ins. Furthermore, tools that are plug-ins by themselves and can be loaded into an integrated development environment (IDE) are well suited. They help to design continuous processes.

The experiences we gained in the project suggest that it is a successful strategy to carefully calculate the effort required for changes to tools. This effort commonly is preferable to the disadvantages of adapting the processes. Besides, customizing tools offer the chance to reflect on the testing processes and optimize them. In the long run, even small changes have great effect. Irregularities caused by hardly changeable tools are likely to cut productivity. Moreover, when tools are not customized or no tools are introduced at all due to the strategy of saving the effort of selecting and adapting them, the company might loose competitiveness. Improving processes and using cutting-edge tools will improve testing and raise the quality of the developed software.

## VI. Conclusion

In this paper we presented results from a project that aimed at finding best practices for software development and especially testing. We described its background, the research approach and the framework used to categorize recommendations. Out of about 30 recommendations found and classified with the framework, we presented five recommendations that make novel contributions to the technical dimension of how testing can be done in enterprises.

Using a modern IDE greatly supports development. It enables testing to focus on finding bugs rather than on eliminating mistakes that entered the code by neglectfulness. Using test case management and a test case database leads to a structured testing process. Moreover, it supports regression testing. Alignment of testing and productive systems prevents many problems that arise due to incompatibilities and scaling issues. Integrating testing and development tools requires continuous governance but increases testing performance and efficiency. Consequently, regression tests can be run much more efficiently. Finally, customizing tools to fit with processes should be preferred over changing processes in order to be able to work with tools.

We found a discrepancy of testing knowledge described in the literature and the reality of testing in enterprises. To give an example: Even practitioners literature such as [1] distinguishes between *black box* and *white box* testing. However, hardly any of the project participants made an explicit distinction like this. Not a single participant had ever heard of *gray box* tests. The above described recommendations might thus be partly found in the literature— but many companies have not implemented them, yet. Apparently, literature is inaccessible for some practitioners, not practically usable in the everyday work, or unknown [30]. This has also been found for organizational aspects of testing [20]. Research progress and testing improvements that were hoped for [14] seem to have reached the industry only partly.

Developing software of high quality is not a mere economic obligation. Neither is it just needed to improve the idea the general public has about software quality. Preventing that software harms humans in any way is an ethical obligation [11]. We thus encourage further research in both organizational areas (i.e. information systems research) and in the technical field (e.g. computer science and formal methods). Moreover, we encourage enterprises to reach a culture of testing instead of perceiving testing as a costly delay in the development process. We therefore propose a structured approach and to keep research bound to cooperation with enterprises.

## VII. Future Work

The project this work is based on is continued in order to evaluate the recommendations found. Future work will contain a discussion of the results with practitioners and probably a quantitative study. Ideally, a study could be done on a national or even global scale. It could not only try to capture the success of the recommendations but check how the literature on software testing is used.

It is without question that a quantitative analysis would perfectly augment the qualitative approach. For example, measuring a return-on-investment (ROI) of the improvements made would be ideal [28]. Without quantitative data it is hard to prove that a recommendation is effective and efficient. Deriving best practices is a first step and was very laborious due to the problematic nature of software testing. Verifying results implemented by companies was identified as a further step. Design science—the research approach of our choice—is incrementally iterative [15]; adding additional rigor and verifying results actually implemented by companies was identified as a further step.

## Acknowledgment

The author would like to thank Herbert Kuchen for discussing the recommendations and for his continuous support and encouragement. Special thanks go to the companies that participated in the project. It would not have been possible without their willingness and help.

## AUTHORS PROFILE

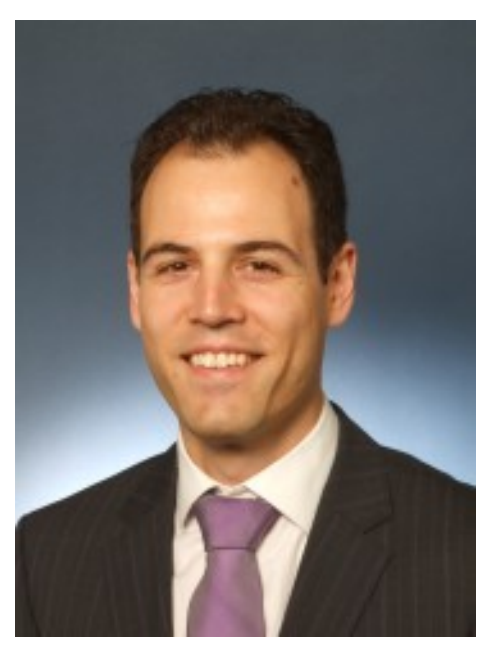

**Tim A. Majchrzak** is a research associate at the Department of Information Systems of the University of Münster, Germany, and the European Research Center for Information Systems (ERCIS). He received a BSc and MSc in Information Systems from the University of Münster and currently finishes his PhD. His research comprises both technical and organizational aspects of software testing. He has also published work on several other interdisciplinary IS topics.